\documentclass[sigconf]{format/acmart}
\usepackage{silence}\SafeMode
\usepackage{enumitem}
\setlist{listparindent=1.25em}
\usepackage[figuresright]{rotating}
\usepackage{tabularx}
\usepackage{balance}
\usepackage{tikz}
\usepackage{pgfplots,pgfplotstable}
\pgfplotsset{compat=1.7}
\usepgfplotslibrary{statistics}
\usepackage{multirow}
\usepackage[normalem]{ulem}
\useunder{\uline}{\ul}{}
\usetikzlibrary{shapes.geometric, arrows, positioning, patterns}
\usepackage{microtype}

\usepackage{newfloat}
\DeclareFloatingEnvironment[
    fileext=loa,
    listname={List of Appendices},
    name=Appendix,
    placement={!ht},
    within=none,
]{floatappendix}

\newcommand\identifying[1]{#1}

\AtBeginDocument{%
  \providecommand\BibTeX{{%
    \normalfont B\kern-0.5em{\scshape i\kern-0.25em b}\kern-0.8em\TeX}}}

\copyrightyear{2026}
\acmYear{2026}
\setcopyright{cc}
\setcctype{by}
\acmConference[SIGCSE TS 2026]{Proceedings of the 57th ACM Technical Symposium on Computer Science Education V.1}{February 18--21, 2026}{St. Louis, MO, USA}
\acmBooktitle{Proceedings of the 57th ACM Technical Symposium on Computer Science Education V.1 (SIGCSE TS 2026), February 18--21, 2026, St. Louis, MO, USA}
\acmPrice{}
\acmDOI{10.1145/3770762.3772529}
\acmISBN{979-8-4007-2256-1/26/02}
\acmConference[arXiv Pre-print]{arXiv Pre-print}{N.D.}{Online}
\acmBooktitle{arXiv Pre-print}
\acmDOI{(arXiv Pre-print)}
\acmISBN{(arXiv Pre-print)}

\begin{document}
\title[The Open Source Resume]{The Open Source Resume: How Open Source Contributions\\ Help Students Demonstrate Alignment with Employer Needs}
\author{Utsab Saha}
\email{usaha@csumb.edu}
\orcid{0009-0004-0458-8697}
\affiliation{
 \institution{Computing Talent Initiative}
 \city{Marina}
 \state{California}
 \country{USA}
}
\additionalaffiliation{
 \institution{California State University, Monterey Bay}
 \city{Marina}
 \state{California}
 \country{USA}
}

\author{Jeffrey D'Andria}
\email{jeffreydandria@gmail.com}
\orcid{0009-0006-1761-0763}
\affiliation{
 \institution{Computing Talent Initiative}
 \city{Marina}
 \state{California}
 \country{USA}
}

\author{Tyler Menezes}
\email{tylermenezes@codeday.org}
\orcid{0000-0002-7975-2533}
\affiliation{
 \institution{CodeDay}
 \city{Seattle}
 \state{Washington}
 \country{USA}
}

\begin{abstract}
Computer science educators are increasingly integrating open sou\-rce contributions into classes to prepare students for higher expectations due to GenAI, and to improve employment outcomes in an increasingly competitive job market. However, little is known about how employers view student open source contributions.

This paper addresses two research questions qualitatively: what traits do employers desire for entry-level hires in 2025, and how can they be demonstrated through open source contributions? It also tests quantitatively the hypothesis that student knowledge of employers' expectations will improve their motivation to work on open source projects.

To answer our qualitative questions, we conducted interviews with US hiring managers. We collaborated with each interviewee to create a ``hiring manager agreement,'' which listed desirable traits and specific ways to demonstrate them through open source, along with a promise to interview some students meeting the criteria. To evaluate our quantitative hypothesis, we surveyed 650 undergraduates attending public universities in the US using an instrument based on expectancy--value theory.

Hiring managers wanted many non-technical traits that are difficult to teach in traditional CS classes, such as initiative. There were many commonalities in how employers wanted to see these traits demonstrated in open source contributions. Viewing hiring manager agreements improved student motivation to contribute to open source projects.

Our findings suggest that open source contributions may help CS undergraduates get hired, but this requires sustained engagement in multiple areas. Educators can motivate students by sharing employer expectations, but further work is required to determine if this changes their behavior.
\end{abstract}



\maketitle

\section{Introduction}

Despite the intrinsic value of education, most students pursuing an undergraduate degree are doing so to improve their job prospects \citep{gedyeStudentsUndergraduateExpectations2004,nortonPerceivedBenefitsUndergraduate2017}, and computer science (CS) students are no exception \citep{helpsStudentExpectationsComputing2005,alshahraniUsingSocialCognitive2018}. One of the most important factors predicting whether a student will obtain a job after graduation is whether they engaged in work-based learning, most commonly internships. However, many students do not participate in such activities because of competing life priorities or because their college is not a ``target'' school for university recruiters. In addition, the job market is becoming increasingly competitive for new graduates. AI coding tools now allow senior developers to quickly accomplish much of the work traditionally given to entry-level developers, so the expectations on new CS graduates are higher than ever.

Many educators are considering open source software contributions as a way to bring more work-based learning into the classroom \citep[e.g.,][]{nascimentoUsingOpenSource2013,hangIndustryMentoringInternship2024,marmorsteinOpenSourceContribution2011,mullerEngagingStudentsOpen2019,choiOpenSourceSoftware2021}. The authors run one such program, in which thousands of students have made their first contributions to open source software in a structured educational experience\identifying{ \citep{parraClosingGapClassrooms2021,narayananScalableApproachSupport2023,menezesOpenSourceInternshipsIndustry2022}}.

The goal of this work is to improve employment outcomes by grounding open source contribution objectives in employer expectations and improving student motivation. Accordingly, this paper addresses two qualitative research questions: \textbf{(RQ1)} What traits do employers desire for entry-level hires? \textbf{(RQ2)} How can students demonstrate those traits through open source work?
Also tested is the following quantitative hypothesis: \textbf{(H1)} Student knowledge of employers' desired traits and demonstrations through open source will improve their motivation to work on open source.
\section{Background}

\subsection{Employers Describe a Need for Skills Acquired in Work-Based Learning}

Skill surveys of industry representatives inform the alignment of class content with relevant industry skills. Such surveys are regularly conducted in CS \citep[e.g.,][]{yahyaMappingGraduateSkills2024, stepanovaHiringCSGraduates2021, scaffidiEmployersNeedsComputer2018, scaffidiSurveyEmployersNeeds2018}. Although many surveys focus exclusively on specific technical skills that can be taught in the classroom, several studies have found that, for CS and other STEM students, employers place a high value on cross-cutting technical skills, the ability to work independently, and ``soft skills'' \citep{scaffidiSurveyEmployersNeeds2018,raynerEmployerPerspectivesCurrent2015,cheangEmployersExpectationsUniversity2023,menezesWhatSkillsCS2023,humeAreWeDeveloping2024,stalhaneWhatCompetenceSoftware2020}. This trend is also evident to students, who report a desire for more career-related training \citep{gedyeStudentsUndergraduateExpectations2004,craigListeningEarlyCareer2018}, and to recent graduates, who report a disconnect between what they learned in college and their first job \citep{begelStrugglesNewCollege2008,kapoorUnderstandingCSUndergraduate2019}.

The skills students learn in the classroom can, as \citet{beaubouefComputerScienceCurriculum2011} put it, ``simply enable [students] to be able to learn what they need to know later on.'' Industry experiences help students develop soft skills and learn to research, experiment, and make decisions independently. For this reason, many undergraduate CS degrees now integrate industry engagement opportunities such as guest speakers, industry mentoring, and case studies. Some programs have even experimented with industry partnerships to provide credit for ``on-the-job'' learning equivalents to classes like data structures, web apps, and databases \citep{carmichaelCurriculumAlignedWorkIntegratedLearning2018}. Internships, however, are the most common form of work-based learning in CS.

It should be no surprise, then, that internships can significantly influence how easily recent graduates transition into the workforce. In fact, whether a student has undertaken an internship is one of the most important variables predicting whether, and how quickly, they have a job after graduation \citep{callananAssessingRoleInternships2004, saltikoffPositiveImplicationsInternships2017, knouseRelationCollegeInternships1999}.

\subsection{Barriers to Internship Access and Open Source as an Alternative}

Unfortunately, internships are scarce. A recent report from the Business--Higher Education Forum found that the supply of internships is insufficient \citep{bhefExpandingInternshipsHarnessing2024}. Of CS undergraduates, only 20\% of rising sophomores and under half of rising juniors/seniors secure internships. By graduation, just 60\% have participated in an internship \citep{kapoorExploringParticipationCS2020, kocClass2014Student2014}. They are also less accessible to lower-income students: most students with a household income over \$150,000 per year graduate with an internship, versus only 35\% of students with a household income under \$100,000 per year \citep{kapoorExploringParticipationCS2020}. There are several reasons why access to internships is difficult for lower-income students:
\begin{itemize}
    \item They may not be able to commit to an internship because they are a primary caregiver or must work a part-time job to pay for school \citep{kapoorBarriersSecuringIndustry2020}.

    \item They may not apply due to lack of confidence \citep{kapoorBarriersSecuringIndustry2020}.

    \item They may not have time to take necessary actions to prepare and apply: successful students spend 3 hours weekly on applications versus 1 hour for unsuccessful ones \citep{kapoorBarriersSecuringIndustry2020,kapoorExploringParticipationCS2020}.

    \item They may attend smaller, more affordable schools that have a harder time capturing the attention of recruiters.
\end{itemize}
Traditional internships are also difficult to integrate into the classroom, and they occur late in students' educational pathways \citep{luceroStructureCharacteristicsSuccessful2021,martincicCombiningRealWorldInternships2009}.

In response to these deficiencies, many colleges have experimented with having students contribute to open source software projects \citep{choiOpenSourceSoftware2021,mullerEngagingStudentsOpen2019,hangIndustryMentoringInternship2024,menezesOpenSourceInternshipsIndustry2022,lovellContributorCatalystPilot2024,lovellScaffoldingStudentSuccess2021,dossantosmontagnerLearningProfessionalSoftware2022,salernoBarriersSelfEfficacyLargeScale2023,hislopStudentReflectionsLearning2020}. Open source software is licensed so that users can read, change, and redistribute its source code\citep{OpenSourceDefinition}. Moreover, anyone is welcome to contribute changes to the original project, where they may be used by others. Popular open source projects include Linux (the operating system kernel), React (the JavaScript library for building user interfaces developed by Meta), and WordPress (a content management system that powers over 40\% of websites). These projects have communities of thousands to hundreds of thousands of contributors, some paid and some who donate their time.

Educators consider many factors when bringing open source into the classroom \citep{nascimentoUsingOpenSource2013}, but few studies have focused on how employers view these open source contributions.

\subsection{Frameworks for Student Motivation}

We considered several frameworks to undergird our investigation into student motivation. All emphasize the role of beliefs and expectations in motivation and career development, though they approach these concepts from different angles.

One of the largest barriers to participation in work-based learning opportunities is low self-efficacy. \citet{hackettSelfefficacyApproachCareer1981} developed a model of career decision self-efficacy that identified four main factors: (1) performance accomplishments, (2) vicarious learning, (3) emotional state, and (4) verbal persuasion.

Social cognitive career theory (SCCT) suggests that individuals are more motivated to undertake activities that might advance their careers and to overcome setbacks when they believe they are competent and have a chance at a positive outcome \citep{lentUnifyingSocialCognitive1994}. SCCT provides a more comprehensive explanation of career development, incorporating social and environmental factors alongside individual beliefs. It also emphasizes the reciprocity among these influences---how experiences shape beliefs, which influence choices and, in turn, create new experiences.

Expectancy‑-value theory (EVT) \citep{wigfieldExpectancyvalueTheory2009} posits that students' achie\-vement related choices are influenced by multiple interconnected factors including social and cultural contexts, affective responses, and individual beliefs about the value of academic tasks. Specifically, EVT examines how students' perceptions of achievement value---encompassing intrinsic value, attainment value, and utility value---along with their assessment of participation costs, collectively shape their engagement in particular academic tasks or activities \citep{ecclesMotivationalBeliefsValues2002,wigfieldDevelopmentCompetenceBeliefs2002,wigfieldExpectancyvalueTheory2009}. EVT focuses on the motivational components of specific tasks or domains, making it particularly useful for understanding moment-to-moment engagement and effort.

We chose EVT because of its wide acceptance in undergraduate educational contexts, and because research suggests that both expectancy and value predict career-related decisions \citep{renningerCambridgeHandbookMotivation2019}.
\section{Methods}

\subsection{Reflexive Interviews with Employers}

For our conversations with employers, we employed reflexive interviewing, a collaborative technique that extends beyond traditional interview methods by engaging both interviewer and interviewee in the co-construction of meaning and understanding. Unlike conventional interviews in which researchers primarily extract information, reflexive interviewing creates a dialogical space where participants collaborate in exploring, elaborating, and interpreting their own experiences and perspectives. The interview process is iterative, allowing for deeper investigation of themes that surface during the conversation \citep{pessoaUsingReflexiveInterviewing2019}.

Technology professionals involved in hiring entry-level software engineers were recruited from program partner companies, meetup attendees, and our personal networks to participate in individual online meetings using a purposive sampling approach. During these one-on-one sessions, participants engaged in interviews exploring the specific traits and competencies they prioritized when evaluating junior developer candidates. Interviews were framed around ``ideal'' candidates. Specifically, we asked the participants to answer the question, ``What would you need to see on a resume so that you would feel like you made a mistake if you didn't interview this person?'' Participants were then asked to provide concrete examples of how these desired traits could be demonstrated through open source contributions made by college students.

During each interview, the insights were synthesized into individualized ``hiring manager agreements'' that captured each participant's desired qualities and evidence criteria. These documents underwent iterative editing as participants reflected on and refined their stated requirements in partnership with the researcher. The iterative nature of this approach allowed for deeper exploration of initially ambiguous or contradictory requirements, with participants able to clarify and elaborate on their expectations through continued dialogue.

Surveys of employers are frequently subject to response bias, especially politeness bias \citep{leeBestAnswerReally2019}, social desirability bias \citep{furnhamResponseBiasSocial1986}, and acquiescence bias \citep{knowlesAcquiescentRespondingSelfReports1997}. In an effort to reduce these biases, we asked each industry professional to e-sign an agreement committing to interviewing an agreed number of students who met their documented requirements. We also expected that this would provide additional motivation for students.

\subsection{Student Motivation Survey}

Student motivation was evaluated through a survey using EVT instruments from two sources---\citet{ecclesMindActorStructure1995} and \citet{flakeMeasuringCostForgotten2015}---but adapted to refer to an open source contribution rather than mathematical skills (as in \citet{ecclesMindActorStructure1995}) or a specific class (\citet{flakeMeasuringCostForgotten2015}). These instruments were selected and modified in prior work by the authors and based on the work of others \citep{olewnikCocurricularEngagementEngineering2023}. The final instrument is presented in Appendix~\ref{tab:motivationSurvey}.

As part of our program, students were already assigned to sections on the basis of time availability; one of these sections was randomly selected as the experimental group. The control group comprised the students in all other sections in the same program in the same year. All students were enrolled in an extracurricular program that included interview preparation, resume advice, and the opportunity to contribute to open source projects \citep{narayananScalableApproachSupport2023}.

Students in the experimental group were shown the hiring manager agreements as part of a regularly scheduled workshop and provided with a link allowing them to read the agreements in detail, before being asked to complete the motivation survey. Responses to the student motivation survey were compared between students in the experiment and control groups. After the survey was completed, all students were provided with equal access to the hiring manager agreements.
\section{Results}
\subsection{Employer Interviews}

\pgfplotstableread[row sep=\\,col sep=&]{
    theme & prevalence \\
    Persistence & 8 \\
    Product Sense & 10 \\
    Curiosity & 9 \\
    Domain Knowl. & 10 \\
    Teamwork & 13 \\
    Initiative & 14 \\
    Problem Solving & 17 \\
}\themesData

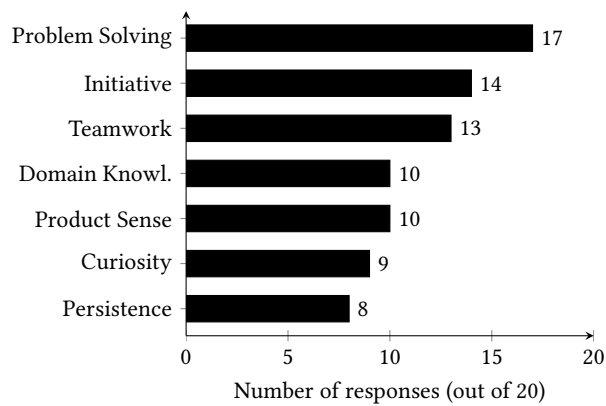
\begin{figure}
    \Description[Bar chart showing theme prevalence in hiring manager agreements]{Bar chart showing theme prevalence in hiring manager agreements. Problem solving and initiative were the most common themes, while teamwork and domain knowledge were seen in about half of agreements. Curiosity and product sense were also included in a small number of agreements.}
    \centering
    \hspace{0.6cm}
    \begin{tikzpicture}
        \begin{axis}[
                xbar,
                symbolic y coords={Persistence,Curiosity,Product Sense,Domain Knowl.,Teamwork,Initiative,Problem Solving},
                ytick=data,
                xmin=0,
                xmax=20,
                xlabel={Number of responses (out of 20)},
                y=0.6cm,
                width=7cm,
                axis lines = left,
                enlarge y limits={0.1},
                nodes near coords={\pgfmathprintnumber\pgfplotspointmeta},
                xticklabel={$\pgfmathprintnumber{\tick}$},
                ]
            \addplot [fill=black] table[x=prevalence,y=theme]{\themesData};
        \end{axis}
    \end{tikzpicture}
    \caption{Prevalence of themes in hiring manager agreements}
    \label{fig:themesChart}
\end{figure}
\begin{table*}
\caption{Themes in hiring manager agreements and corresponding examples of open source demonstration}
\label{tab:themes}
\begin{tabular}{p{2cm}p{15cm}}
\toprule
\textbf{Theme} & \textbf{Open Source Evidence Examples} \\ \midrule
\begin{tabular}[c]{@{}l@{}}Problem\\ Solving\end{tabular} & \begin{tabular}[c]{@{}p{14cm}}``10 issues solved; a sufficient variety of types of issues they tackle; issues touch on different parts of the technology stack or solve different kinds of problems.; at least one issue should be labeled as medium or hard.''\vspace{0.15cm} \\ ``6 issues solved. The issues should be reasonably challenging (required a change across multiple modules)''\end{tabular} \\ \midrule
Initiative & \begin{tabular}[c]{@{}p{14cm}}``1 example of solving a problem for the project that wasn’t directly asked for (e.g., they open and solve an issue themselves, initiate a new feature, or improve the project’s technical documentation); 2 examples of helping onboard someone into the project (e.g., answering questions on the project’s forum).''\vspace{0.15cm} \\ ``2 examples of a PR where the definition of `done' goes beyond the bare minimum of what is asked for in the original issue. Do they think about details that are not explicitly asked for (e.g., writing unit tests or updating documentation)? Do they consider other aspects of the product, like performance or security?''\end{tabular} \\ \midrule
\begin{tabular}[c]{@{}l@{}}Teamwork/\\ Collaboration\end{tabular} & \begin{tabular}[c]{@{}p{14cm}}``2 examples of comments on a PR that is not their own, where their feedback is accepted by the other person. 2 examples of a pull request where the candidate received critical feedback and they incorporated the feedback.''\vspace{0.15cm} \\ ``2 examples of thorough documentation on a pull request: document the process of how they initiated the new issue/feature. [...] Even better if the issue has comments throughout the timeline of the Github thread.\end{tabular} \\ \midrule
\begin{tabular}[c]{@{}l@{}}Domain\\ Knowledge\end{tabular} & \begin{tabular}[c]{@{}p{14cm}}``Experience with tools that SRE’s might use: Data Dog, Grafana, Splunk, Prometheus''\vspace{0.15cm} \\ ``Jupyter notebooks (API’s), Docker/Kubernetes, Node, Unix Shell''\end{tabular} \\ \midrule
Curiosity & \begin{tabular}[c]{@{}p{14cm}}``2 examples of finding and filing new bugs on the project. 1 example of fixing a complex bug (getting to the root cause), where they had to explore multiple branches of possibilities. The solution is explained in a blog post, video walkthrough, or detailed pull request description.''\vspace{0.15cm} \\ ``2 examples of a PR where they consider multiple approaches to solving the problem. Describe the multiple possibilities in the pull request. Bonus points for making a case for your best recommendation.''\end{tabular} \\ \midrule
\begin{tabular}[c]{@{}l@{}}Product\\ Sense\end{tabular} & \begin{tabular}[c]{@{}p{14cm}}``You can state your opinion on what new features should be added to an open source project you are involved in and explain why your recommendation would solve a problem for the customers/users of the product''\vspace{0.15cm} \\ ``You can clearly articulate the purpose of the project. You can name specific projects/organizations who use it and/or you can describe the primary use cases for a typical end user.'\end{tabular} \\ \midrule
Persistence & \begin{tabular}[c]{@{}p{14cm}}``Time to completion averages to 3 months or less''\vspace{0.15cm} \\ ``1 example of a long-running PR where you follow through to the end. Even better if the issue has comments throughout the timeline of the PR.''\end{tabular} \\ \bottomrule
\end{tabular}
\end{table*}

A total of 20 interviews were conducted with hiring managers at technology companies. While theoretical saturation was achieved after 12 interviews, 8 additional interviews were completed to strengthen the robustness of the findings and address reviewer recommendations. Among the interviewees, 5 were C-level executives (3 at mid-sized companies, 2 at small startups); 2 were senior leadership (all mid-sized companies); and the remaining 13 were mid-to-senior-level managers (8 Fortune 500, 5 mid-sized).

Hiring managers were willing to interview various numbers of students. The smallest commitment was 3 interviews (2 hiring managers) and the largest was 10 (1 hiring manager); all the remaining hiring managers (17) were willing to interview 5 students.

When asked about the traits they desired, 7 common themes were identified (Figure~\ref{fig:themesChart}, Table~\ref{tab:themes}):  \textbf{problem solving} (ability to solve complex problems without hand-holding), \textbf{initiative} (solving problems without being asked), \textbf{teamwork/collaboration} (communication and coachability), \textbf{domain knowledge} (specific technologies used at the company), \textbf{curiosity} (learning for the sake of it), \textbf{product sense} (understanding the product and users), and \textbf{persistence} (working until a problem is solved).

\subsection{Student Motivation}

We recruited 650 students to complete the motivation survey; 591 (the control group) did not view the hiring manager agreements, and 59 (the experimental group) did. Students were recruited from juniors and seniors at state-funded universities and community and technical colleges in the US.

Data normality was assessed using the Shapiro--Wilk test for each construct within both groups. The normality assumption was satisfied for all constructs ($p > 0.05$). Data for some constructs were found to violate equal-variance assumptions, so Welch $t$ tests were performed to compare means across groups for each construct. To control for multiple comparisons, $p$ values were adjusted using the Holm--Bonferroni method.

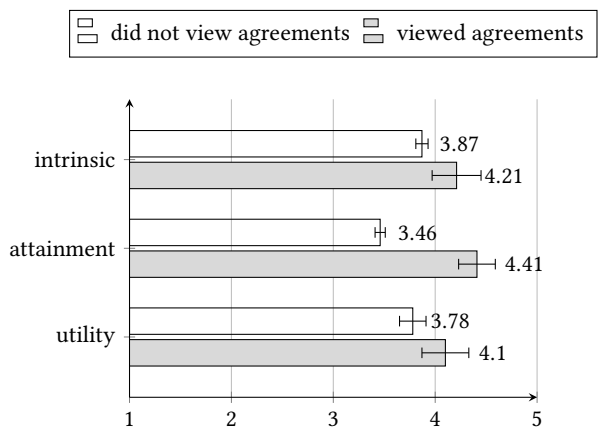
\begin{figure}
    \centering
    \hspace{0.6cm}
    \begin{tikzpicture}
        \begin{axis}[
                xbar,
                symbolic y coords={utility,attainment,intrinsic},
                ytick=data,
                xmin=1,
                xmax=5,
                y=1.18cm,
                width=7cm,
                axis lines = left,
                enlarge y limits={0.34},
                nodes near coords={\pgfmathprintnumber\pgfplotspointmeta},
                nodes near coords align={shift={(0.55cm,0)}},
                xtick={1,2,3,4,5},
                xticklabel style={align=center},
                xmajorgrids=true,
                reverse legend,
                legend style={at={(0.5,1.3)},
                    anchor=north,legend columns=-1
                    ,inner sep=3pt,column sep=3pt},
                ]
            \addplot [fill=gray!30]
            plot [nodes near coords align={shift={(0.63cm,0)}}, error bars/.cd, x dir=both, x explicit]
            table[x=agreement,y=construct,x error=agreementErr] {\valueData};
            \addplot [black]
            plot [nodes near coords align={shift={(0.5cm,0)}}, error bars/.cd, x dir=both, x explicit]
            table[x=baseline,y=construct,x error=baselineErr]{\valueData};
            \legend{viewed agreements,did not view agreements}
        \end{axis}
    \end{tikzpicture}
    \caption{Comparison of motivational beliefs about value constructs by hiring manager agreement viewing status: higher indicates more motivated (99\% C.I., $n$ = 650)}    \Description[Bar chart showing motivational beliefs about value constructs]{Bar chart showing motivational beliefs about value constructs by agreement viewing status, with 99\% confidence intervals. Intrinsic and attainment beliefs increased beyond confidence interval bounds after viewing agreements, while utility increased within the bounds.}
    \label{fig:value}
\end{figure}

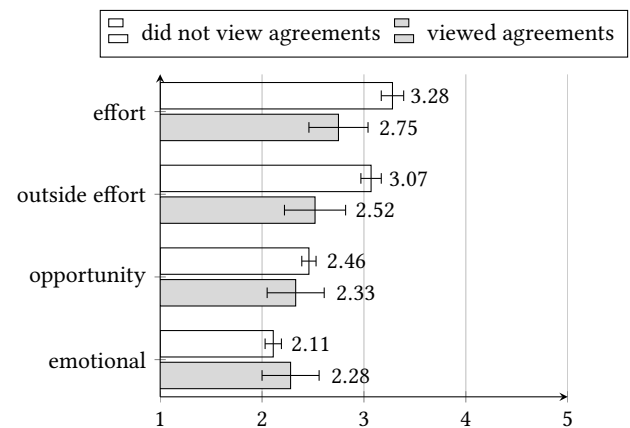
\begin{figure}
    \centering
    \hspace{0.6cm}
    \begin{tikzpicture}
        \begin{axis}[
                xbar,
                symbolic y coords={emotional,opportunity,outside effort,effort},
                ytick=data,
                xmin=1,
                xmax=5,
                y=1.1cm,
                width=7cm,
                axis lines = left,
                enlarge y limits={0.15},
                nodes near coords={\pgfmathprintnumber\pgfplotspointmeta},
                nodes near coords,
                xtick={1,2,3,4,5},
                xticklabel style={align=center},
                xmajorgrids=true,
                reverse legend,
                legend style={at={(0.5,1.2)},
                    anchor=north,legend columns=-1
                    ,inner sep=3pt,column sep=3pt},
                ]
            \addplot [fill=gray!30]
            plot [nodes near coords align={shift={(0.8cm,0)}}, error bars/.cd, x dir=both, x explicit]
            table[x=agreement,y=construct,x error=agreementErr] {\costData};
            \addplot [black]
            plot [nodes near coords align={shift={(0.5cm,0)}}, error bars/.cd, x dir=both, x explicit]
            table[x=baseline,y=construct,x error=baselineErr]{\costData};
            \legend{viewed agreements,did not view agreements}
        \end{axis}
    \end{tikzpicture}
    \caption{Comparison of motivational beliefs about cost constructs by hiring manager agreement viewing status: lower indicates more motivated (99\% C.I., $n$ = 650)}
    \label{fig:cost}
    \Description[Bar chart showing motivational beliefs about cost constructs]{Bar chart showing motivational beliefs about cost constructs by agreement viewing status, with 99\% confidence intervals. Effort and outside effort beliefs decreased beyond confidence interval bounds after viewing agreements, while opportunity cost decreased within bounds, and emotional cost increased within bounds.}
\end{figure}

The results (Figures \ref{fig:value} and~\ref{fig:cost} and Table~\ref{tab:stats}) revealed statistically significant differences between students who had and had not viewed the hiring manager agreements across all constructs ($p < 0.01$ to $p < 0.0001$) except Opportunity and Emotional Costs. Effect size analyses indicated a small practical effect for Utility Value; a moderate effect for Intrinsic Value, Effort Cost, and Outside Effort Cost; and a large effect for Attainment Value ($d = 2.17$). All significant differences favored increased motivation among students who viewed the agreements.

\pgfplotstableread[row sep=\\,col sep=&]{
    construct   & baseline  & agreement & baselineErr   & agreementErr\\
    intrinsic   & 3.87      & 4.21	    & 0.06          & 0.24 \\
    attainment  & 3.46      & 4.41      & 0.05          & 0.18 \\
    utility     & 3.78      & 4.10      & 0.13          & 0.23 \\
}\valueData

\pgfplotstableread[row sep=\\,col sep=&]{
    construct       & baseline  & agreement & baselineErr   & agreementErr \\
    effort          & 3.28      & 2.75      & 0.11          & 0.29 \\
    outside effort  & 3.07      & 2.52      & 0.10          & 0.30  \\
    opportunity     & 2.46      & 2.33      & 0.07          & 0.28 \\
    emotional       & 2.11      & 2.28      & 0.08          & 0.28 \\
}\costData

\begin{table}
\caption{Comparison of motivational beliefs by hiring manager agreement viewing status}

\label{tab:stats}
\begin{tabular}{lcccc}
\toprule
\textbf{} & \textbf{\begin{tabular}[c]{@{}c@{}}Not Viewed\\ Mean (SD)\end{tabular}} & \textbf{\begin{tabular}[c]{@{}c@{}}Viewed\\ Mean (SD)\end{tabular}} & \textbf{\begin{tabular}[c]{@{}l@{}}$t$\ test\end{tabular}} & \textbf{\begin{tabular}[c]{@{}c@{}}\small{Cohen's}\\ $d$\end{tabular}} \\
\midrule
\multicolumn{5}{c}{\textit{\textbf{Value (higher indicates more motivated)}}} \\
Intrinsic       & \(3.87\) (0.55)   & \(4.21\) (0.70)   & *    & \(0.61\)\\
Attainment      & \(3.46\) (0.43)   & \(4.40\) (0.51)   & ***  & \(2.17\)\\
Utility         & \(3.78\) (1.22)   & \(4.10\) (0.67)   & *    & \(0.27\)\\
\multicolumn{5}{c}{\textit{\textbf{Cost (lower indicates more motivated)}}} \\
Effort          & \(3.28\) (1.05)   & \(2.75\) (0.84)   & **    & \(-0.52\)\\
Outside Effort  & \(3.06\) (0.90)   & \(2.52\) (0.88)   & **    & \(-0.61\)\\
Opportunity     & \(2.46\) (0.68)   & \(2.33\) (0.81)   & n.s.  & n.s. \\
Emotional       & \(2.12\) (0.74)   & \(2.28\) (0.80)   & n.s.  & n.s.\\
\bottomrule
\end{tabular}

{\footnotesize * $p < 0.01$, **  $p<0.001$, ***  $p<0.0001$, n.s. = not statistically significant}
\end{table}
\section{Discussion}

\subsection{RQ1: Employers' Desired Traits}

Of all the traits identified as important by hiring managers, only \textit{domain knowledge} (identified in 10 interviews) was easy to teach in lecture-focused courses. This demonstrates the importance of providing work-based learning opportunities.

The emphasis on \textit{initiative} (14 interviews), \textit{product sense} (10 interviews), and \textit{curiosity} (9 interviews) reflects a fundamental tension in CS education. Traditional coursework is largely assignment driven: students complete prescribed tasks with clear requirements and deadlines. However, employers want people who identify problems and solve them proactively and who are intrinsically motivated to grow their skills over time. This suggests that educators should create more open-ended learning experiences in which students must define their own problems to solve, rather than just implementing solutions to given specifications.

Two highly rated traits---\textit{problem solving} (17 interviews) and \textit{teamwork/collaboration} (13 interviews)---are, in the authors' experience, rarely addressed in a rigorous way. Given their importance among employers, educators should consider adding courses on these subjects.

\subsection{RQ2: Demonstration Through Open Source}

Although hiring managers had different thresholds and specifics, there were many commonalities in how they wanted students to demonstrate key traits through open source contributions. Employers wanted 6--10 issues solved at a variety of difficulty levels or in different applications or parts of an application (\textit{problem solving}), for students to solve problems that were not directly asked for (\textit{initiative}), and for students to help with documentation and participate in code reviews (\textit{technical communication}).

There was less consistency in how employers wanted students to demonstrate \textit{domain knowledge}, \textit{curiosity}, \textit{product sense}, or \textit{persistence}. However, based on our conversations we suspect that many hiring managers would still be enthusiastic about candidates who demonstrate these traits in other ways.

The desired \textit{domain knowledge} was strongly tied to the specific team for which the individual was hiring. CS degrees that offer tracks/concentration/focuses may have an advantage here.

\subsection{H1: Impact on Motivation of Students' Knowledge of Employer Expectations}

Students who viewed the hiring manager agreements were more motivated by the value of the open source contributions, particularly the Attainment Value. Students who viewed the hiring manager agreements also thought the costs of contributing were lower. (Cronbach's $\alpha$ was low for Utility Value, suggesting that it may be unreliable.) This finding is perhaps the most actionable for educators. It suggests that explicit discussion of hiring criteria with employers could improve student engagement with educational activities.
\section{Limitations and Future Work}

Employer commitments are a positive step, but ultimately we hope for jobs. This paper leaves open whether hiring managers actually interview students meeting their requirements and whether those interviews lead to job offers. The qualitative results would also benefit from a much larger sample of employers.

On student perception, this study only looked at whether the students who viewed hiring manager agreements demonstrated a change in motivation; it did not measure behavior change. It may also be biased by the fact that the students surveyed were all enrolled in an extracurricular program that included the opportunity to make an open source contribution. Such students may have higher motivation to contribute to open source and to take actions to obtain a career than the general undergraduate population.
\begin{floatappendix*}
\caption{Student motivation survey}
\label{tab:motivationSurvey}
\begin{tabular}{llp{11.85cm}c}
\toprule
\multicolumn{2}{l}{\textbf{Construct}} & \textbf{Question (5-point Likert)} & \textbf{Cronbach's $\alpha$} \\ \midrule
\multicolumn{1}{l}{\multirow{7}{*}{\begin{tabular}[c]{@{}l@{}}\textbf{Value}\\ \textbf{\cite{ecclesMindActorStructure1995}}\end{tabular}}} & \multirow{2}{*}{\textbf{Intrinsic}} & In general, I think I would find working on open source projects... & \multirow{2}{*}{0.78} \\ \cmidrule{3-3}
\multicolumn{1}{l}{} &  & How much do you think you would like working on open source projects? &  \\ \cmidrule{2-4} 
\multicolumn{1}{l}{} & \multirow{3}{*}{\textbf{Attainment}} & Is the amount of effort that it would take to do well in open source contributions worthwhile to you? & \multirow{3}{*}{0.73} \\ \cmidrule{3-3}
\multicolumn{1}{l}{} &  & I feel that participating in open source activities is \rule{4em}{0.5pt} to my being a competent engineer. &  \\ \cmidrule{3-3}
\multicolumn{1}{l}{} &  & How important would it be to you to do well in open source work? &  \\ \cmidrule{2-4} 
\multicolumn{1}{l}{} & \multirow{2}{*}{\textbf{Utility}} & How useful do you think it would be to participate in open source work for what you want to do after you graduate? & \multirow{2}{*}{0.58} \\ \cmidrule{3-3}
\multicolumn{1}{l}{} &  & How useful do you think it would be to participate in open source work for your daily life outside school? &  \\ \midrule
\multicolumn{1}{l}{\multirow{10}{*}{ \begin{tabular}[c]{@{}l@{}}\textbf{Cost}\\ \textbf{\cite{flakeMeasuringCostForgotten2015}}\end{tabular}}} & \multirow{3}{*}{\textbf{Effort}} & Participating in open source would take up too much time. & \multirow{3}{*}{0.76} \\ \cmidrule{3-3}
\multicolumn{1}{l}{} &  & I would have to put too much energy into open source work. &  \\ \cmidrule{3-3}
\multicolumn{1}{l}{} &  & Participating in open source work would require too much effort. &  \\ \cmidrule{2-4} 
\multicolumn{1}{l}{} & \multirow{2}{*}{\begin{tabular}[c]{@{}l@{}}\textbf{Outside}\\ \textbf{Effort}\end{tabular}} & I have so many other commitments that I can't put forth the effort needed for open source work. & \multirow{2}{*}{0.7} \\ \cmidrule{3-3}
\multicolumn{1}{l}{} &  & Because of all the other demands on my time, I don't have enough time for open source work. &  \\ \cmidrule{2-4} 
\multicolumn{1}{l}{} & \multirow{2}{*}{\textbf{Opportunity}} & Working on open source work would cause me to miss out on too many things I care about. & \multirow{2}{*}{0.72} \\ \cmidrule{3-3}
\multicolumn{1}{l}{} &  & I wouldn't be able to spend as much time doing the other things that I like to do if I did open source work. &  \\ \cmidrule{2-4} 
\multicolumn{1}{l}{} & \multirow{3}{*}{\textbf{Emotional}} & I would be concerned about being embarassed if my work in open source is inferior to that of my peers. & \multirow{3}{*}{0.74} \\ \cmidrule{3-3}
\multicolumn{1}{l}{} &  & Working on open source work would be emotionally draining. &  \\ \cmidrule{3-3}
\multicolumn{1}{l}{} &  & Working on open source work would be too frustrating. &  \\ \bottomrule
\end{tabular}
\end{floatappendix*}
\section{Conclusions}

The ``bar'' for an ideal software engineering graduate is high, and, with the exception of domain knowledge, the traits that employers desire are difficult to teach through conventional assignments. Although the minimum bar to get hired is likely lower, market data show that unemployment rates are rising among recent CS graduates as AI is becoming better positioned to perform entry-level work. Educators should prepare students for the higher bar.

Bringing open source software into the classroom does seem to be a viable solution, and employer expectations for open source work are theoretically achievable, but the depth and type of engagement they would like to see are more extensive than we have observed in classroom-integrated work-based learning so far. Educators taking the open source route may wish to consider deepening their engagement. The good news is that students who learn about employer expectations seem willing to take on some open source work as an extracurricular activity, though it remains to be seen whether this leads to measurable behavior change.

\section*{Online Resources}

Supplemental files, including anonymized versions of hiring manager agreements, are available on CodeDay's website at:\\ \url{https://doi.org/10.69924/dkr1ysunc72wyzsdgtlrykdn} \cite{sahaSupplementalFilesOpen2025}

\begin{acks}
Thank you to Lola Egherman for reviewing a draft of this paper. This material is based upon work supported by the National Science Foundation under Grant No. 2347311.
\end{acks}

\bibliographystyle{format/ACM-Reference-Format}
\balance
\bibliography{arxiv}

\end{document}